\documentclass[preprint,superscriptaddress,showpacs]{revtex4-1}

\usepackage[utf8]{inputenc} % set input encoding (not needed with XeLaTeX)

\usepackage{amsfonts}
\usepackage{amsbsy}
\usepackage{ulem}
\usepackage[margin=2cm]{geometry}

\usepackage{xcolor}
\usepackage{mathrsfs}
\usepackage{hyperref}
\usepackage{lineno}
\usepackage{graphicx} % support the \includegraphics command and options

\usepackage{booktabs} % for much better looking tables
\usepackage{array} % for better arrays (eg matrices) in maths
\usepackage{paralist} % very flexible & customisable lists (eg. enumerate/itemize, etc.)
\usepackage{verbatim} % adds environment for commenting out blocks of text & for better verbatim
\usepackage{subfigure} % make it possible to include more than one captioned figure/table in a single float
\usepackage[utf8]{inputenc}
\usepackage[T1]{fontenc}

\usepackage{fancyhdr} % This should be set AFTER setting up the page geometry
\begin{document}
%\linenumbers
\title{Entropic bounds between two thermal equilibrium states}
\author{Julio A. \surname{L\'opez-Sald\'ivar}}
\email{Corresponding author: \, julio.lopez@nucleares.unam.mx}
\affiliation{Instituto de Ciencias Nucleares, Universidad Nacional Autónoma de Mexico, Apdo. Postal 70-543, 04510, CDMX, Mexico}
\affiliation{Moscow Institute of Physics and Technology (State University), Institutskii per. 9, Dolgoprudnyi, Moscow Region 141700, Russia}
\author{ Octavio \surname{Casta\~nos}}
\email{ocasta@nucleares.unam.mx}
\affiliation{Instituto de Ciencias Nucleares, Universidad Nacional Autónoma de Mexico, Apdo. Postal 70-543, 04510, CDMX, Mexico}
\author{ Margarita A. \surname{Man'ko}}
\email{mmanko@sci.lebedev.ru}
\affiliation{Lebedev Physical Institute, Russian Academy of Sciences, Leninskii Prospect 53, Moscow 119991, Russia}
\author{ Vladimir I. \surname{Man'ko}}
\email{manko@sci.lebedev.ru}
\affiliation{Moscow Institute of Physics and Technology (State University), Institutskii per. 9, Dolgoprudnyi, Moscow Region 141700, Russia}
\affiliation{Lebedev Physical Institute, Russian Academy of Sciences, Leninskii Prospect 53, Moscow 119991, Russia}
\affiliation{Tomsk State University, Department of Physics, Lenin Avenue 36, Tomsk 634050, Russia}

\begin{abstract}
The positivity conditions of the relative entropy between two
thermal equilibrium states $\hat{\rho}_1$ and $\hat{\rho}_2$ are
used to obtain upper and lower bounds for the subtraction of their
entropies, the Helmholtz potential and the Gibbs potential of the
two systems. These limits are expressed in terms of the mean values
of the Hamiltonians, number operator, and temperature of
the different systems. In particular, we discuss these limits for
molecules which can be represented in terms of the Franck--Condon
coefficients. We emphasize the case where the Hamiltonians belong to
the same system at two different times $t$ and $t'$. Finally, these
bounds are obtained for a general qubit system and for the harmonic
oscillator with a time dependent frequency at two different times.
\end{abstract}

\maketitle

\section{Introduction}
The states of quantum systems are described either by vectors
$\vert \psi\rangle$ in a Hilbert space ${\cal H}$~\cite{Dirac-book}
(pure states) and the corresponding wave functions $\psi(x)=\langle
x \vert \psi\rangle$~\cite{Schroed1926,*Schroed19261} or by the density operators
$\hat\rho$ acting in the Hilbert space~\cite{Landau27,vonNeumann27}
(mixed states).  These states are associated to Hamiltonian  system interactions with certain environments or
external sources. The systems can consist of a constant number of
particles or, due to the interactions, can have a
varying number of particles. In view of this, there are mixed
states with density operators in equilibrium, depending on such
physical parameters as the temperatures and chemical potentials.

In quantum mechanics, one can find various characteristics of
arbitrary pure and mixed states in terms of the von~Neumann~\cite{vonNeumannbook}, Tsallis~\cite{Tsallis}, and
R\'enyi entropies~\cite{Renyi}, as well as known equalities and
inequalities; see, for example,~\cite{Lieb,*Lieb2,Ruskai,*Ruskaie,Araki}. On the other hand, the correlations of a
system with an external source can be of such a form that it
preserves the purity of the states but the system
Hamiltonian depends on time; this means that the system energy
changes due to the interchange with the external source, and this
change is described by the time dependence of the Hamiltonian
parameters.

The variations of the parameters may be either very slow or very fast
with respect to the relaxation time of the system. In the case of
a very fast change in the Hamiltonian parameters (instantaneous rate of change), the
studied state, being either pure or mixed, just after the perturbation
continues to be the same as it was before due to inertia. Thus, if the system was in the pure state with a
given energy level, the wave function just after the perturbation does not change in
spite of the fact that the Hamiltonian is modified. Similarly if the system
was in a thermal equilibrium mixed state with the density operator
$\hat{\rho}$, this operator is the same just after the
instant Hamiltonian parameter variation, though the Hamiltonian itself
is different.

Our generic approach is to study the bounds for the state
characteristics making use of the relative entropy between two thermal states,
concentrating the applications of the inequalities to the thermal
equilibrium states and their possible changes. This is related to
the developments of the studies of states associated with quantum
thermodynamics. In fact, in recent years the analysis of the
thermodynamic properties of the information (quantum and classical)
has been the subject of several works
\cite{huber,parrondo,maruyama,esposito,brandao,skrzypczyk,faist,strasberg,campisi,mancino}.
In particular, the fundamental thermodynamical aspects of
information as the second law, the Landauer principle and the
Maxwell's demon have been studied \cite{plenio,maruyama,esposito}.  In relation with the fundamental aspects of statistical mechanics, we stress that in~\cite{Popescu2006} a general canonical principle has been proposed without using temporal or ensemble averages, which however can be easily connected to the standard statistical mechanics. How
entanglement and coherence can be used to generate work has been the
subject of \cite{Brandao2008,funo13,ren,korzekwa16}. The
problem of how the thermodynamic quantities as the internal energy,
the entropy, and the Helmholtz potential behave as a system approach
equilibrium has been of interest. These investigations have led to
the definition of different inequalities regarding these quantities
\cite{esposito,figueroa,julio}.

 In  the previous works \cite{figueroa, julio}, we have analyzed the comparison between an arbitrary state with the density matrix $\hat{\rho}$  and a thermal equilibrium state $\hat{\sigma}=e^{-\hat{H}/T}/{\rm Tr}(e^{-\hat{H}/T})$ making use of the Tsallis and  von Neumann relative entropies. This comparison was made in specific for a qubit system and a Gaussian state resulting in an inequality that relates the entropy of $\hat{\rho}$, the mean value ${\rm Tr}(\hat{\rho} \hat{H})$, and the partition function of the system $Z(\hat{H},T)={\rm Tr}( e^{-\hat{H}/T})$. The bounds for physical characteristics as the energy or entropy of quantum states play an important role since they determine the specific states which correspond on the extreme situation where the equality between the bound an the physical quantity of interest are equal. In our work \cite{figueroa} it was shown that the distance, given the relative entropy expression, between the arbitrary state with density matrix $\hat{\rho}$ and the canonical thermal equilibrium state with Hamiltonian $\hat{H}$ provides the bound for the sum of the energy and the entropy (in dimensionless variables). Exactly on this bound the canonical Boltzmanian density matrix is realized as it was point out also in \cite{nouvo}. The observation that the physical state of thermal equilibrium is related with the bound gives the motivation to study other bounds in quantum thermodynamics. In this work we study, using the relative entropy as a distance between the quantum states, the bounds for differences of entropies and free energies associated with states corresponding to different Hamiltonians and temperatures. Such bounds give the possibility to study the specific states which appear when the Hamiltonians depend on time. Specifically we are interested in how the system behaves when a sudden change in these parameters is done. Such situation takes place if the duration of the parameter change is smaller than the relaxation time of the system, e.g., in molecular spectroscopy such regime is associated with the Franck-Condon factors which are used to describe the vibronic structure of electronic lines in molecules if the transition takes place between the pure energy level states. We point out that the results from this research can be of importance in the field of  quantum information thermodynamics.

In this work, we obtain new upper and lower limits of the difference
of the entropy, the Helmholtz and Gibbs potentials between two
different thermal equilibrium density matrices $\hat{\rho}_1$ and
$\hat{\rho}_2$. Although these two states can be not related, we make
special emphasis in the case where they describe the same system at
two different times $t$ and $t'$. As, in principle, the initial and
final states may not have the same purity, one can think that the
initial state with number operator $\hat{N}_1$ is in contact with
an external source at temperature $T_1$, whose interaction yields an
effective Hamiltonian $\hat{H}_1$ over the system. At some point, a
change in the interaction $\hat{H}_1\rightarrow \hat{H}_2$, the number  operator $\hat{N}_1 \rightarrow \hat{N}_2$, and the
temperatures $T_1 \rightarrow T_2$ is done by the energy or particle
transfer between the system and the external source, changing the
thermodynamic properties of the system. It is important to stress
that the expressions obtained can be applied to any type of change
done to the system, i.e., if these changes are either quasistatic or
not.

As examples, these bounds are studied for a general qubit system and
the harmonic oscillator with a time-dependent frequency.

\section{Bounds between two systems interchanging energy}
First, we discuss the case where the system is represented by the
Hamiltonian $\hat{H}$ and the parameter $T$, and may
interact with an external source only through energy exchanges. As
it is known, the description of such a system can be done using the
canonical ensemble. In this representation, any state given by the
density operator $\hat{\sigma}=e^{-\hat{H}/T}/{\rm
Tr}(e^{-\hat{H}/T})$ has the von Neumann entropy $S=-{\rm
Tr}(\hat{\sigma}\ln\hat{\sigma})$. This entropy can also be
expressed as $S=\langle
\hat{H} \rangle/T +\ln Z(\hat{H},T)$, where the quantity
$Z(\hat{H},T)={\rm Tr}(e^{-\hat{H}/T})$ is called the partition
function while the parameter $T$ is the temperature. In this case,
the operator $\hat{\sigma}$ describes a thermal equilibrium state
(in a unit system where $\hbar=k=1$).

As we compare an arbitrary nonthermal equilibrium state $\hat{\rho}$
with $\hat{\sigma}$ using the nonnegative relative entropy
\cite{nielsen} ${\rm Tr}(\hat{\rho}(\ln \hat{\rho}-\ln
\hat{\sigma}))$, one can notice that the entropy of the nonequilibrium
system must satisfy the inequality $S < {\rm
Tr}\,(\hat{\rho}\hat{H}) /T + \ln Z(\hat{H},T)$, which can be used
to distinguish the equilibrium state from the nonequilibrium one
\cite{figueroa,julio}. Also, we point out that another inequality can be defined when the operator $\hat{H}$ and the temperature $T$ are replaced by an arbitrary observable $\hat{\mathcal{O}}$ and parameter $\lambda$, respectively, i.e., by doing the replacement $\hat{\sigma}\rightarrow e^{-\hat{\mathcal{O}}/\lambda}/{\rm Tr}(e^{-\hat{\mathcal{O}}/\lambda})$. Later on this idea will be used to find bounds for a grand canonical ensemble.

We consider two thermal equilibrium states described by the
Hamiltonians and temperatures (or arbitrary parameters) ($\hat H_1$,
$T_1$) and ($\hat H_2$, $T_2$), respectively,
\begin{equation}
\hat{\rho}_1=\frac{e^{-\beta_1 \hat{H}_1}}{\textrm{Tr}
\,(e^{-\beta_1 \hat{H}_1})}, \quad \hat{\rho}_2=\frac{e^{-\beta_2
\hat{H}_2}}{\textrm{Tr} (e^{-\beta_2 \hat{H}_2})} \, ,
\label{rhos}
\end{equation}
with $\beta_1=1/T_1$ and $\beta_2=1/T_2$. The difference of their
entropies is given by the following expression:
\begin{equation}
S(\hat{H}_2, T_2)-S(\hat{H}_1, T_1)=\frac{1}{T_2}{\rm
Tr}\,(\hat{\rho}_2 \hat{H}_2)-\frac{1}{T_1}{\rm Tr}(\hat{\rho}_1
\hat{H}_1)+\ln\left(\frac{Z(\hat{H}_2,T_2)}{Z(\hat{H}_1,T_1)}\right)
\, .
\label{deltasa}
\end{equation}
This quantity can be evaluated if either the mean value of the
Hamiltonians and the temperatures or the partition functions of both
systems are known (as the mean value of the Hamiltonian can be
obtained by differentiating the logarithm of the partition function
${\rm Tr}(\hat{\rho}\hat{H})=-\partial \ln Z(\hat{H},T)/\partial
\beta$). On the other hand, it can be shown that, in view of the relative
entropy, upper and lower bounds for the difference of the entropies
$S(\hat{H}_2, T_2)-S(\hat{H}_1, T_1)$ between the two thermal
equilibrium states can be obtained. To demonstrate this, the
positivity conditions $\textrm{Tr}(\hat{\rho}_1 \ln \hat{\rho}_1-
\hat{\rho}_1 \ln \hat{\rho}_2 ) \geq 0$ and
$\textrm{Tr}(\hat{\rho}_2 \ln \hat{\rho}_2- \hat{\rho}_2 \ln
\hat{\rho}_1 ) \geq 0$ are used. From these, the bounds for
$S(\hat{H}_2, T_2)-S(\hat{H}_1, T_1)$ can be written as the
following inequality:
\begin{equation}
\frac{1}{T_2}\left( E(\hat{H}_2, T_2) -\frac{\textrm{Tr}\,
(e^{-\beta_1 \hat{H}_1} \hat{H}_2)}{Z( \hat{H}_1, T_1)} \right)
\leq S(\hat{H}_2, T_2)-S(\hat{H}_1, T_1) \leq \frac{1}{T_1}
\left(\frac{\textrm{Tr}(e^{-\beta_2 \hat{H}_2} \hat{H}_1)}
{Z(\hat{H}_2, T_2)} - E(\hat{H}_1, T_1)\right) \, ,
\label{deltas}
\end{equation}
where $E(\hat{H},T)={\rm Tr}\,(e^{-\beta
\hat{H}}\hat{H})/Z(\hat{H},T)$ is the mean value of the Hamiltonian.
Adding and subtracting $E(\hat{H}_1,T_1)$ and $E(\hat{H}_2,T_2)$ to
the left- and right-hand sides of the previous expression,
respectively, we obtain the following result:
\begin{equation}
\frac{1}{T_2}\left( \Delta E - \frac{{\rm Tr}\,(e^{-\beta_1 \hat{H}_1}
\Delta \hat{H})}{Z (\hat{H}_1,T_1)}\right) \leq \Delta S \leq
\frac{1}{T_1} \left( \Delta E - \frac{{\rm Tr}\,(e^{-\beta_2
\hat{H}_2}\Delta \hat{H})}{Z(\hat{H}_2,T_2)} \right) \, ,
\label{diffe}
\end{equation}
with $\Delta E=E (\hat{H}_2,T_2)-E(\hat{H}_1,T_1)$, $\Delta \hat{H}
=\hat{H}_2-\hat{H}_1$, and $\Delta
S=S(\hat{H}_2,T_2)-S(\hat{H}_1,T_1)$. It is worth mentioning that
the limits for the difference of the entropies are related to the
mean values of the complementary Hamiltonians of each system. When
both density matrices $\hat{\rho}_1$ and $\hat{\rho}_2$ belong to
the same Hilbert space, i.e., when the Hamiltonians $\hat{H}_1$ and
$\hat{H}_2$ are related by a transform that may be not unitary. The term ${\rm Tr}(e^{-\beta_1 \hat{H}_1}
\hat{H}_2)/Z(\hat{H}_1,T_1)$  can be interpreted as the mean value
of the Hamiltonian after the change $\hat{H}_1\rightarrow
\hat{H}_2$, when the change is sudden and the system has no time to
adapt. In this case, the state of the system $\hat{\rho}_1$ remains unchanged,
e.g.,  when the relaxation time of the system is larger compared
with the time when the change of the Hamiltonian occurs. This
behavior is due to the fact that, after changing the Hamiltonian,
the state determined by $\hat{H}_1$ and $T_1$ present an inertia
that prevents it from change very quickly as stated by the adiabatic
theorem of quantum mechanics. The other mean value ${\rm
Tr}(e^{-\beta_2
\hat{H}_2} \hat{H}_1)/Z(\hat{H}_2,T_2)$ is the mean value of the
Hamiltonian, when the system undergoes the change $\hat{H}_2
\rightarrow \hat{H}_1$ and can be interpreted as a reversibility
term. Also as the relative entropy between the two thermal states
$\hat{\rho}_1$ and $\hat{\rho}_2$ measures the distance between the
two states, it can be used to compare the different Hamiltonians
$\hat{H}_1$ and $\hat{H}_2$ which define the two systems. It is also
worth clarifying that the limits for the difference $\Delta S$ are
only valid if the initial and final states are of thermal
equilibrium, e.g., in the following situation: initially the system
is kept in thermal equilibrium at $T_1$ with Hamiltonian $\hat H_1$
for all times $t < t_1$; at $t=t_1$, a change in the temperature and
the interaction Hamiltonian is done until a certain time $t=t_2$.
After this, the system is kept at temperature $T_2$, and interaction
Hamiltonian $\hat H_2$ until it finally achieves thermal
equilibrium. When these conditions are satisfied, the difference of the entropy (or free energy) between the two equilibrium states can be approximated using only the mean value of the Hamiltonian corresponding to the times just before and after an abrupt change in the conditions of the system is done. This implies that measuring the change on the mean value of the Hamiltonian before and after the change one can have a quick estimate of the difference of the thermodynamic quantities even before the systems reach equilibrium.

When the Hamiltonians $\hat{H}_{1,2}$ are written in terms of the kinetic and potential operators $\hat{H}_j=\hat{K}_j+\hat{V}_j$ ($j=1,2$), the bounds for the difference of the entropy can be expressed as: $(\langle \hat{K}_2 \rangle_2+\langle \hat{V}_ 2\rangle_2-\langle \hat{K}_2 \rangle_1-\langle \hat{V}_2 \rangle_1)/T_2\leq \Delta S \leq (\langle \hat{K}_1 \rangle_2+\langle \hat{V}_1 \rangle_2-\langle \hat{K}_1 \rangle_1-\langle \hat{V}_1 \rangle_1)/T_1$ , with $\langle \hat{\mathcal{O}}_j \rangle_k={\rm Tr}(\hat{\rho}_k \hat{\mathcal{O}_j})$. In a situation where the kinetic energy does not change, the limits can be expressed as the difference of the mean values of the potential operators: $(\langle \hat{V}_2 \rangle_2-\langle \hat{V}_2 \rangle_1)/T_2 \leq \Delta S \leq (\langle \hat{V}_1 \rangle_2-\langle \hat{V}_1 \rangle_1)/T_1$.

Furthermore, in view of (\ref{deltasa}) and~(\ref{deltas}), we can
obtain bounds for the function $\ln
(Z(\hat{H}_2,T_2)/Z(\hat{H}_1,T_1))$ as follows:
\begin{equation}
{\rm Tr}\left(\hat{\rho}_1\left(\frac{\hat{H}_1}{T_1}-\frac{\hat{H}_2}
{T_2}\right)\right) \leq \ln \left(\frac{Z(\hat{H}_2,T_2)}{Z(\hat{H}_1,T_1)}
\right) \leq {\rm Tr}\left(\hat{\rho}_2\left(\frac{\hat{H}_1}{T_1}
-\frac{\hat{H}_2}{T_2}\right)\right) \, .
\end{equation}
As the logarithm of the partition function is related to the
Helmholtz potential $F(\hat{H},T)=-T \ln (Z(\hat{H},T))$, the
previous equation allow us to obtain limits for the difference of
the Helmholtz potential of both systems
\begin{equation}
{\rm Tr}\left(\hat{\rho}_1\left(\frac{\hat{H}_1}{T_1}-\frac{\hat{H}_2}
{T_2}\right)\right) \leq \frac{F(\hat{H}_1,T_1)}{T_1}-\frac{F(\hat{H}_2,T_2)}
{T_2} \leq {\rm Tr}\left(\hat{\rho}_2\left(\frac{\hat{H}_1}{T_1}
-\frac{\hat{H}_2}{T_2}\right)\right) \, .
\label{helm}
\end{equation}
These limits as well as the ones for the entropy depend only on the
mean values of the Hamiltonians and the parameters $T_1$ and $T_2$.

When the Hamiltonian operators are written in terms of their
eigenvalues and eigenvectors, i.e.,
\[
\hat{H}_1=\sum_{j}\epsilon_j \vert \epsilon_j \rangle \langle \epsilon_j \vert \,
, \quad \hat{H}_2=\sum_{l}\varepsilon_l \vert \varepsilon_l \rangle \langle \varepsilon_l \vert \, ,
\]
the two density matrices $\hat{\rho}_1$ and $\hat{\rho}_2$ can be
expressed as $\hat{\rho}_1=\sum_j \mathcal{P}_j(T_1) \vert
\epsilon_j \rangle \langle \epsilon_j \vert $ and
$\hat{\rho}_2=\sum_l \mathscr{P}_l(T_2) \vert \varepsilon_l \rangle
\langle \varepsilon_l \vert $, where
$\mathcal{P}_j(T_1)=e^{-\epsilon_j/T_1}/\sum_{j'}e^{-\epsilon_{j'}/T_1}$
and
$\mathscr{P}_l(T_2)=e^{-\varepsilon_l/T_2}/\sum_{l'}e^{-\varepsilon_{l'}/T_2}$
are the probabilities associated to the states $\vert \epsilon_j
\rangle \langle \epsilon_j \vert$ and $\vert \varepsilon_l \rangle
\langle \varepsilon_l \vert$, respectively.

The mean values of $\hat{H}_1$ and $\hat{H}_2$ are equal to ${\rm
Tr}\,(\hat{\rho_1} \hat{H}_1)=\sum_j \mathcal{P}_j (T_1)
\epsilon_j$ and ${\rm Tr(\hat{\rho_2} \hat{H}_2)}=\sum_l
\mathscr{P}_l(T_2)
\varepsilon_l$, while the mean values of the Hamiltonian, using the
complementary system states, read
\begin{equation}
{\rm Tr}(\hat{\rho}_1 \hat{H}_2)=\sum_{j,l} \mathcal{P}_j (T_1)
\varepsilon_l k_{jl} \, , \quad {\rm Tr}(\hat{\rho}_2 \hat{H}_1)= \sum_{j,l}
\mathscr{P}_j (T_2) \epsilon_l k_{lj} \, ,
\end{equation}
where the matrix elements $k_{jl}=\vert \langle \epsilon_j \vert
\varepsilon_l \rangle \vert^2$ are known as the Franck--Condon
factors when the states represent two electronic states in a
molecular system. These factors have been calculated and simulated
for several electronic transitions in molecules, e.g. a compilation of
these factors for the hydrogen molecule ${\rm H}_2$ can be seen
in~\cite{fantz}, and methods to obtain vibronic transition profiles
in molecules have been studied in \cite{doktorov,huh}. Finally, the
previous expressions for the mean value of the Hamiltonians are used
to obtain the following bounds for the difference of the entropy
\begin{equation}
\frac{1}{T_2}\left(\sum_j \mathscr{P}_j (T_2) \varepsilon_j -\sum_{jl}
\mathcal{P}_j (T_1)\varepsilon_l k_{jl}\right)\leq \Delta S \leq
\frac{1}{T_1}\left(\sum_{jl} \mathscr{P}_j(T_2) \epsilon_l k_{lj}
-\sum_j \mathcal{P}_j(T_1) \epsilon_j \right) \, .
\end{equation}
In addition, using the same arguments, the difference of the
Helmholtz potential is
\begin{equation}
\sum_{j} \mathcal{P}_j \epsilon_j/T_1-\sum_{j l} \mathcal{P}_j (T_1)
\varepsilon_l k_{jl}/T_2 \leq \frac{F(\hat{H}_1,T_1)}{T_1}-
\frac{F(\hat{H}_2,T_2)}{T_2} \leq \sum_{jl} \mathscr{P}_j (T_2)
\epsilon_j k_{jl}/T_1 -\sum_j \mathscr{P}_j (T_2) \varepsilon_j/T_2 \, .
\label{fcondon}
\end{equation}

In the case where the two Hamiltonians have the same spectrum
($k_{ij}=\delta_{ij}$), $\Delta S$ will have as bounds the
difference of mean value of the energy corresponding to the two
different temperatures:
\begin{eqnarray}
\frac{1}{T_2}\left( \sum_j (\mathscr{P}_j (T_2)-\mathcal{P}_j (T_1))
\varepsilon_j\right) \leq \Delta S \leq \frac{1}{T_1}\left( \sum_j
(\mathscr{P}_j (T_2)-\mathcal{P}_j (T_1))\epsilon_j\right) \, ,\nonumber \\
\sum_{j}\mathcal{P}_j(T_1) (\epsilon_j/T_1- \varepsilon_j/T_2) \leq
\frac{F(\hat{H}_1,T_1)}{T_1}-\frac{F(\hat{H}_2,T_2)}{T_2} \leq
\sum_j \mathscr{P}_j (T_2) (\epsilon_j /T_1- \varepsilon_j /T_2) \, .
\end{eqnarray}

The inequalities given in Eqs.~(\ref{diffe}),~(\ref{helm}),
and~(\ref{fcondon}) allow us to study the behavior of a system that
experience a sudden change, even if the period of time for this is
very small compared with the relaxation time of the system. In those
cases, the bounds of $\Delta S$ or the Helmholtz potential must be
larger compared with a small change over time. Also these boundaries
can be used as an approximation for $\Delta S$ or $
F(\hat{H}_1,T_1)/T_1-F(\hat{H}_2,T_2)/T_2 $ and have the convenience
to only depend on the mean values of the Hamiltonians opposed to the
analytic expressions whose also depend on the partition function.

\section{Bounds between two systems interchanging energy and \\
particles}

It is possible to obtain an analogous expression for the bounds of
the entropy on a system that interacts interchanging energy and
particles with an external source. In order to describe this kind of
systems, it is necessary to use the grand canonical ensemble in which
a thermal equilibrium state is given by the following density
matrix:
\[
\hat{\sigma}=\frac{e^{\beta(\mu \hat{N}-\hat{H})}}{{\rm Tr}\,
(e^{\beta(\mu \hat{N}-\hat{H})})} \, ,
\]
where $\mu$ is the chemical potential and $\hat{N}$ is the number operator of the different energy levels of the system. When
using the positivity condition of the relative entropy between an
arbitrary state given by the density matrix $\hat{\rho}$ and the
equilibrium matrix $\hat{\sigma}$, the following new
inequality for the von Neumann entropy $S=-{\rm Tr}(\hat{\rho}\ln
\hat{\rho})$ is obtained
\begin{equation}
S\leq \ln (\mathscr{Z}(\hat{H},\hat{N},T,\mu))-\frac{1}{T}{\rm
Tr}\,(\hat{\rho}(\mu \hat{N}-\hat{H}))) \, ,
\end{equation}
where $\mathscr{Z}(\hat{H}, \hat{N},T,\mu)={\rm Tr}\,(e^{\beta(\mu
\hat{N}-\hat{H})})$ is the grand partition function. As in the
canonical ensemble, the equality of the previous expression only
occurs when the system is in thermal equilibrium. Therefore, this
new inequality can be used to distinguish between equilibrium and
nonequilibrium states in a general system.

As in the canonical case, the comparison between two different
equilibrium states
\[
\hat{\rho}_1 = \frac{e^{\beta_1 (\mu_1 \hat{N}_1-\hat{H}_1)}}{{\rm Tr}\,
(e^{\beta_1 (\mu_1 \hat{N}_1-\hat{H}_1)})} \, , \quad \hat{\rho}_2 =
\frac{e^{\beta_2 (\mu_2 \hat{N}_2-\hat{H}_2)}}{{\rm Tr}\,
(e^{\beta_2 (\mu_2 \hat{N}_2-\hat{H}_2)})}
\]
can be performed. The von Neummann relative entropy conditions ${\rm
Tr}\,(\hat{\rho}_1 \ln \hat{\rho}_1-\hat{\rho}_1 \ln
\hat{\rho}_2)
\geq 0$ and ${\rm Tr}\,(\hat{\rho}_2 \ln \hat{\rho}_2-\hat{\rho}_2 \ln
\hat{\rho}_1)\geq 0$ give rise to the following limits for the
difference of the entropies $S(\hat{H}_2,T_2)$ and
$S(\hat{H}_1,T_1)$:
\begin{eqnarray*}
\frac{1}{T_2}\left(  \frac{{\rm Tr}\,(e^{\beta_1 (\mu_1 \hat{N}_1
-\hat{H}_1)}(\mu_2 \hat{N}_2-\hat{H}_2))}{\mathscr{Z}(\hat{H}_1,
\hat{N}_1,T_1,\mu_1)}- G(\hat{H}_2,\hat{N}_2,T_2,\mu_2)+{\rm Tr}\,(\hat{\rho}_2
\hat{H}_2)\right)   \leq\Delta S \leq \\  
\leq  \frac{1}{T_1}\left( G(\hat{H}_1,
\hat{N}_1,T_1,\mu_1)-{\rm Tr}(\hat{\rho}_1 \hat{H}_1) -\frac{{\rm Tr}\,
(e^{\beta_2 (\mu_2 \hat{N}_2-\hat{H}_2)}(\mu_1
\hat{N}_1-\hat{H}_1))}
{\mathscr{Z}(\hat{H}_2,\hat{N}_2,T_2,\mu_2)}\right),
\end{eqnarray*}
where $G(\hat{H},\hat{N},T,\mu)={\rm Tr}\,(\hat{\rho}\hat{H})+T(\ln
(\mathscr{Z}(\hat{H},\hat{N},T,\mu))-S(\hat{H},\hat{N},T,\mu))$ is the Gibbs
potential. The term ${\rm Tr}\,(\hat{\rho}_1\hat{N}_2)$ is
interpreted as the new mean value of the number operator when
the system undergoes the sudden changes $\hat{N}_1 \rightarrow
\hat{N}_2$ and $\hat{H}_1 \rightarrow \hat{H}_2$. These sudden
changes are thought to be faster than the relaxation time of the
system, hence the state does not change. We notice that these limits
can be used to estimate the entropy change of a system, having the
convenience of depending on mean values of observable quantities;
also it can be used to detect changes in the Hamiltonian, the number of particles or the temperature of the system even in the
occurrence of a very fast transform. The previous inequality can
also be used to obtain bounds for the logarithm of the ratio of the
grand partition functions
\begin{eqnarray*}
\frac{1}{T_1}({\rm Tr}\,(\hat{\rho}_1 \hat{H}_1)- G(\hat{H}_1,
\hat{N}_1,T_1,\mu_1))+\frac{1}{T_2}{\rm Tr}\,(\hat{\rho}_1(\mu_2 \hat{N}_2
-\hat{H}_2)) \leq \ln \left( \frac{\mathscr{Z}(\hat{H}_2,
\hat{N}_2,T_2,\mu_2)}{\mathscr{Z}(\hat{H}_1,\hat{N}_1,T_1,\mu_1)}\right) \leq \\
\leq \frac{1}{T_2}(G(\hat{H}_2,\hat{N}_2,T_2,\mu_2)-{\rm Tr}\,(\hat{\rho}_2
\hat{H}_2))-\frac{1}{T_1}{\rm Tr}(\hat{\rho}_2(\mu_1 \hat{N}_1-\hat{H}_1))\, .
\end{eqnarray*}

To see some applications of these inequalities, we present briefly
the case of a general qubit system.

\section{Qubit system}
In recent years, the study of qudit systems has been of great
importance due to its use in quantum information, in particular, the
study of the qubit system and its interaction with different
environments. In this section, we present the entropic inequalities
between two different qubit systems.

The study discussed in the previous section is used to present the
entropy bounds between two different qubit systems generated by the
Hamiltonians $\hat{H}_1$ and $\hat{H}_2$ and described by the
density matrices $\hat{\rho}_1$ and $\hat{\rho}_2$ of
Eq.~(\ref{rhos}), which are expressed in terms of the Bloch vectors
for the corresponding systems, i.e.,
\begin{equation}
\hat{H}_1= \frac{1}{2} \left( \begin{array}{cc}
h_0+h_3 & h_1 - i h_2 \\
h_1+ i h_2 & h_0- h_3
\end{array} \right), \quad
\hat{H}_2= \frac{1}{2} \left( \begin{array}{cc}
\mathfrak{h}_0+\mathfrak{h}_3 & \mathfrak{h}_1 - i \mathfrak{h}_2 \\
\mathfrak{h}_1+ i \mathfrak{h}_2 & \mathfrak{h}_0- \mathfrak{h}_3
\end{array} \right),
\end{equation}
where $\mathbf{h}=(h_1,h_2,h_3)$ and $\boldsymbol{\mathfrak{h}}=
(\mathfrak{h}_1,\mathfrak{h}_2, \mathfrak{h}_3)$ are the Bloch
vectors of $\hat{H}_1$ and $\hat{H}_2$, respectively, while $h_0$ and $\mathfrak{h}_0$ are the traces of the Hamiltonians. Here we have use the Bloch representation of the states although some
other representations can be used as the ones
presented in \cite{dodonov1, Chernega2017}.  

The entropy for each system is a function of the norm of the Bloch
vector of the Hamiltonian and the temperature; it reads
\begin{eqnarray}
S(\vert \mathbf{h} \vert, T_1)=\frac{1}{2} \left( \ln 2+ \ln
\left(1+\cosh \left( \frac{\vert \mathbf{h} \vert}{T_1}\right)\right)
-\frac{\vert \mathbf{h} \vert}{T_1} \tanh\left( \frac{\vert \mathbf{h}
\vert}{2 T_1}\right) \right) \, . \nonumber
\end{eqnarray}
The corresponding expression for $S(\vert \boldsymbol{\mathfrak{h}}
\vert, T_2)$ can be obtained making the substitution  $\mathbf{h}
\rightarrow \mathfrak{h}$ and $T_1 \rightarrow T_2$. The mean value
of the Hamiltonian is
\begin{eqnarray}
E(\hat{H}_1, T_1)= \frac{1}{2}\left( h_0 -\vert \mathbf{h} \vert \tanh
\left( \frac{ \vert \mathbf{h} \vert}{2 T_1} \right) \right), \end{eqnarray}
while the mean value of the Hamiltonian $\hat{H}_2$ seen in the
complementary system $\hat{\rho}_1$ is
\begin{eqnarray}
\textrm{Tr}\,(\hat{\rho}_1 \hat{H}_2)= \frac{1}{2 \vert \mathbf{h} \vert}
\left( \vert \mathbf{h} \vert \mathfrak{h}_0 -  \mathbf{h} \cdot
\boldsymbol{\mathfrak{h}} \tanh \left( \frac{ \vert \mathbf{h} \vert}
{2 T_1}\right) \right) \, . \nonumber
\end{eqnarray}
Using these expressions, one can write the upper and lower bounds
for the difference of the entropy as
\begin{eqnarray}
\frac{\vert \boldsymbol{\mathfrak{h}}\vert}{2 T_2} \left( \cos \theta \tanh
\left( \frac{ \vert \mathbf{h} \vert}{2 T_1}\right)-\tanh
\left( \frac{ \vert \boldsymbol{\mathfrak{h}} \vert}{2 T_2}\right)\right)
\leq S(\vert \boldsymbol{\mathfrak{h}} \vert, T_2) - S(\vert \mathbf{h}
\vert, T_1) \leq \nonumber \\
\leq \frac{\vert \mathbf{h} \vert}{2 T_1} \left( \tanh
\left( \frac{ \vert \mathbf{h}\vert}{2 T_1}\right)-\cos \theta \tanh
\left( \frac{ \vert \boldsymbol{\mathfrak{h}}\vert}{2 T_2}\right)\right)
\label{eq:qubit} ,
\end{eqnarray}
where $\theta$ is the angle between the two Bloch vectors $\mathbf{h}$ and $\boldsymbol{\mathfrak{h}}$.

By differentiating the upper and lower bounds with respect to
$\theta$, we see that the lower bound has a minimum value when
$\theta=\pi$ and a maximum when $\theta=0$, while the upper bound
has a minimum at $\theta=0$ and a maximum at $\theta=\pi$. From
these extreme values it is possible to see that the limits are
closer to the exact value of $\Delta S$ when the upper bound has a
minimum, and the lower bound has a maximum ($\theta=0$), and present
the largest difference comparing with the exact value when the upper
bound has a maximum and the lower bound has a minimum
($\theta=\pi$). Then one can conclude that the Hamiltonians
$\hat{H}_1$ and $\hat{H}_2$ which give rise to thermal equilibrium
states (at the same temperature $T$), with the same von Neumann
entropy ($\Delta S =0$), are the ones that have parallel Bloch
vectors and the ones, which give rise to a maximum difference of
the entropy between them, have antiparallel Bloch vectors.

%Figure 1
\begin{figure}
\centering
\subfigure[]
{\includegraphics[scale=0.25]{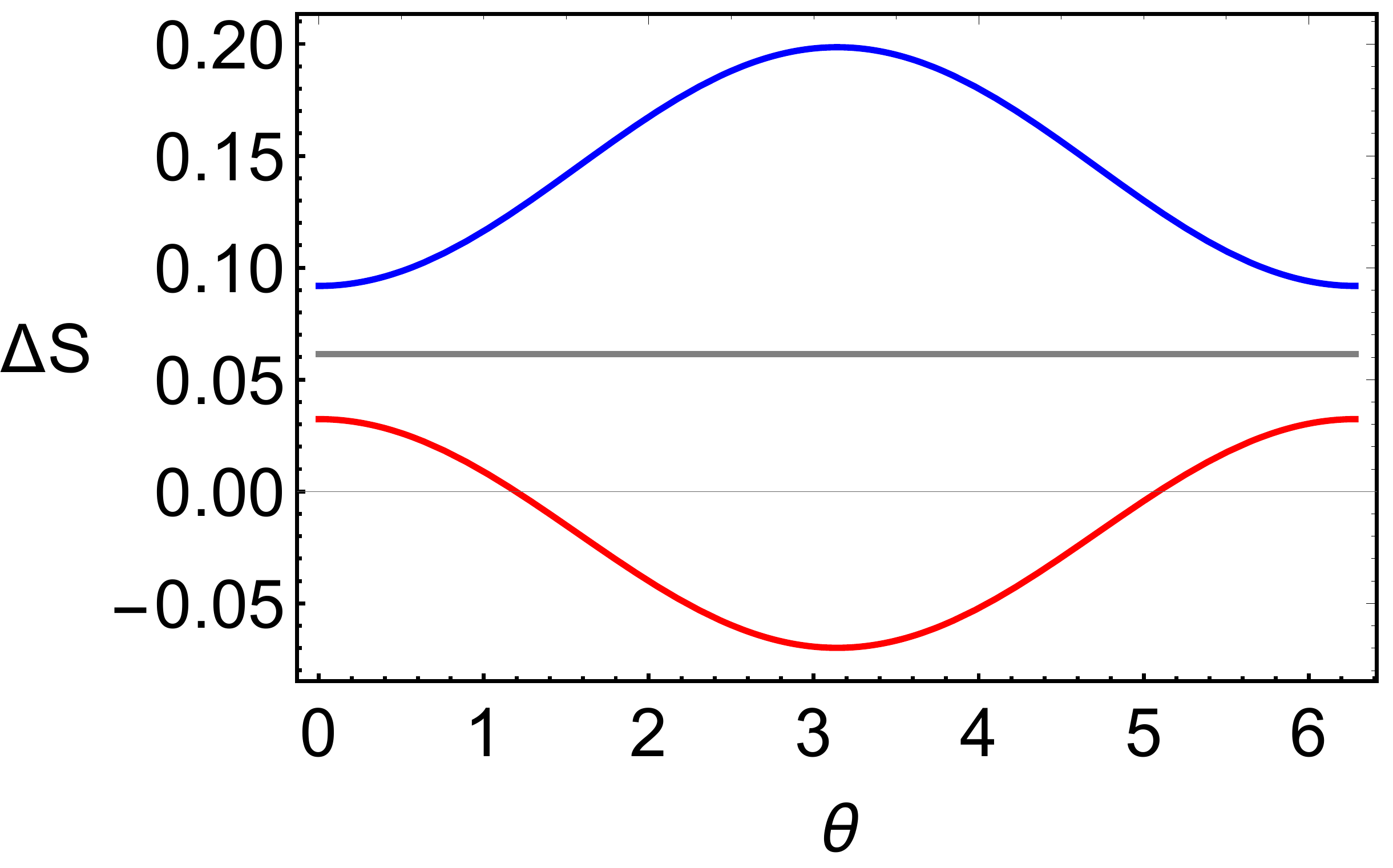} \label{fig:qubit:a}} \
\subfigure[]
{\includegraphics[scale=0.25]{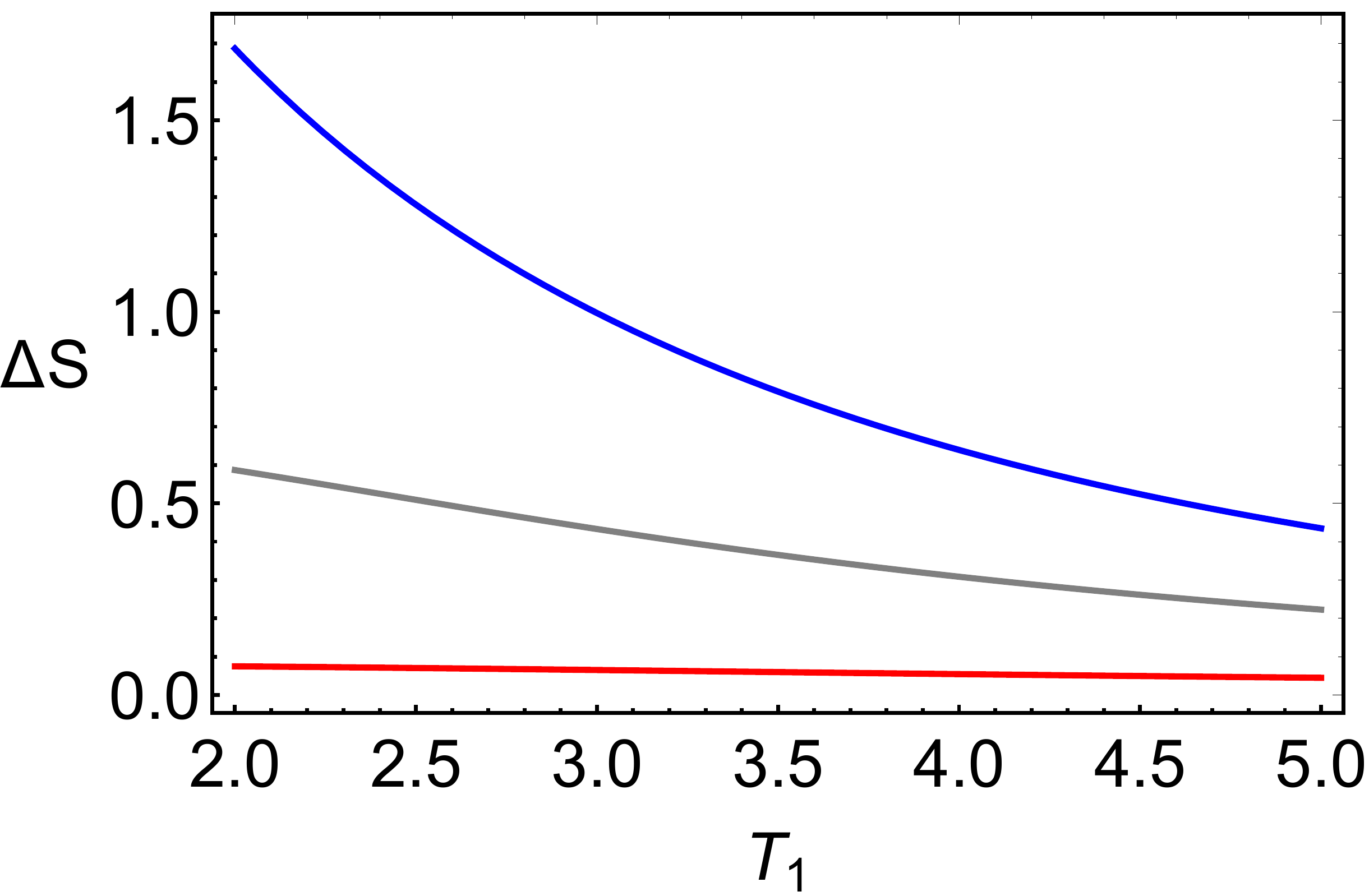} \label{fig:qubit:b}}
\caption{ Upper (blue) and lower (red) bounds for the difference
of entropies $\Delta S$~(a) as a function of  the angle between the
Bloch vectors $\theta$ between two thermal equilibrium states with
temperature $T_1=10$ and (b) as a function of temperature $T_1$ with
fixed $\theta=\pi/4$. In both cases, we assume $T_2=15$, $\vert
h\vert=\sqrt{61}$, and $\vert \mathfrak{h} \vert=\sqrt{17}$. The
gray curves correspond to the analytic result.}
\end{figure}

In fig.~\ref{fig:qubit:a}, the upper and lower bounds of $\Delta S$
are shown for a qubit system as a function of the angle $\theta$
between the Bloch vectors $ \mathbf{h} $ and
$\boldsymbol{\mathfrak{h}}$. One can see that these limits have a
minimum value when the Bloch vectors are parallel and a maximum when
they are antiparallel as previously discussed. In
fig.~\ref{fig:qubit:b}, the plot of $\Delta S$ is shown as a
function of temperature $T_1$ with fixed $T_2$. Here, one can see
that the difference between the bounds and the analytic curve in
gray goes down as the temperature increases. This is due to the fact
that, as the temperature increases the density matrices
$\hat{\rho}_1$ and $\hat{\rho}_2$ become more and more similar to
the most mixed state $\mathbf{I}/2$ (e.g., a spin system where the
probability of being up and down is the same), independently of the
Hamiltonians $\hat{H}_1$ and $\hat{H}_2$.

\section{Harmonic oscillator with a time-dependent frequency}

The time dependent harmonic oscillator \cite{husimi} has been a
paradigmatic model in quantum mechanics \cite{dodonov}. This kind of
oscillator may have exact solutions and can be used to obtain
statistical properties of the electromagnetic field as antibunching
and squeezing \cite{mandal}. Additionally, a scheme to calculate the
Franck--Condon factors for two one-dimensional harmonic
oscillators have been studied in \cite{octavio}.

In this section, the study of the bounds for the harmonic oscillator
with a time dependent frequency is presented making use of the time dependent invariant operators of the Hamiltonian, although other different methods can be used e.g. using the Heisenberg operators at two different times. The Hamiltonians are
given at two different times, i.e., $\hat{H}_1=\hat{H}(t)$ and $\hat{H}_2=H(t')$. The Hamiltonian of the system  is
\begin{equation}
\hat{H}(t)=\frac{1}{2}(\hat{p}^2+ \omega^2 (t) \hat{q}^2) \, ,
\end{equation}
which has a the time dependent invariant operator $\hat{A}(t)=i(\epsilon (t)
\hat{p}-\dot{\epsilon}(t)\hat{q})/\sqrt{2}$,  with $\epsilon(t)$
being a solution of the classical equation $\ddot{\epsilon}(t)+\omega^2 (t)
\epsilon(t)=0$ with the initial conditions $\epsilon (0)=1$, $\dot{\epsilon}(0)=i$.
This operator satisfies the bosonic commutation relation
$[\hat{A}(t),\hat{A}^\dagger (t)] =1$ implying the property
$\dot{\epsilon}(t) \epsilon^* (t) -\epsilon(t) \dot{\epsilon}^*
(t)=2i$.

From this it is possible to define the integral of motion operator
$\hat{A}^\dagger(t) \hat{A}(t)$, which has the eigenfunctions
\begin{equation}
\phi_n (x,t)= \left(\frac{\epsilon^* (t)}{2 \, \epsilon(t)}\right)^{n/2}
\frac{e^{i\frac{\dot{\epsilon}(t)}{2 \, \epsilon(t)}x^2}}{\sqrt{n!\,
\epsilon(t) \pi^{1/2}}} \,  H_n \left( \frac{x}{\vert \epsilon(t) \vert}\right) \, .
\end{equation}
These eigenfunctions form a complete orthonormal set at any time,
i.e., $\int dx \, \phi_m^* (x,t) \phi_n(x,t)=\delta_{nm}$,  and
satisfy also the closure condition $\sum_n \phi_n^*(x, t)
\phi_n(x',t)=\delta(x-x')$. In order to obtain the upper and lower
bounds of $\Delta S$, one needs to calculate the mean values of the
Hamiltonian at any time and the value $\textrm{Tr}(e^{- \beta
\hat{H}(t')} \hat{H}(t))$. To perform these calculations, it is convenient
to write the Hamiltonian at a time $t'$ in terms of the operators at time $t$, i.e.,
\begin{equation}
\hat{H}(t')=\alpha (t,t') \hat{K}_- (t)+ \alpha^* (t,t') \hat{K}_+ (t)
+ \gamma (t,t') \hat{K}_0 (t) \, ,
\label{hal}
\end{equation}
where we have defined the functions
\[
\alpha (t,t')= \frac{1}{2}(\dot{\epsilon}^{*2}(t)+\omega^2 (t')
\epsilon^{*2}(t)), \quad \gamma (t, t')=\vert \dot{\epsilon}(t)
\vert^2+\omega^2 (t') \vert \epsilon (t)\vert^2 \, ,
\]
which satisfy the relation $4 \omega^2 (t')=\gamma^2 (t,t')- 4 \vert \alpha(t,t')\vert^2$, and the operators $\hat{K}_-(t)=\hat{A}^2 (t)/2$, $\hat{K}_+ (t)=\hat{A}^{\dagger
2} (t)/2 $, and $\hat{K}_0 (t)=(\hat{A}(t)\hat{A}^\dagger
(t)+\hat{A}^\dagger (t)\hat{A}(t))/4$ are the generators of the
SU(1,1) group. Thus, the mean value of $\hat{H}$ at time $t$ in the
$t'$ state is
\[
\frac{1}{Z(\hat{H}(t'),T)}{\rm Tr}(e^{-\beta \hat{H}(t')}\hat{H}(t))=\frac{1}{Z(\hat{H}(t'),T)}\sum_{n=0}^\infty
\langle n,t \vert e^{-\beta \hat{H}(t')}\hat{H}(t) \vert n, t \rangle \, ,
\]
which gives the result (see appendix A)
\begin{equation}
\frac{1}{Z(\hat{H}(t'),T)}{\rm Tr}(e^{-\beta \hat{H}(t')}\hat{H}(t))
= \frac{\omega^2(t)+\omega^2(t')}{4 \omega (t')} \coth \left( \frac{\omega (t')}{2T}\right) \, .
\label{prom}
\end{equation}
This expression only depends on the frequencies at the different times and the temperature and not in the classical solutions $\epsilon (t)$ and $\epsilon (t')$ which greatly simplify their calculation. One can notice that the mean value at time $t$ of the Hamiltonian
${\rm Tr}(e^{-\beta \hat{H}(t)}\hat{H}(t))/Z(\hat{H}(t),T)$ can be
obtained from the previous expression making $t'=t$ that provides
the result
\begin{equation}
\frac{1}{Z(\hat{H}(t),T)}{\rm Tr}(e^{-\beta \hat{H}(t)}\hat{H}(t))
=\frac{\omega (t)}{2} \coth \left( \frac{\omega (t)}{2 T}\right) \, ,
\end{equation}

Finally from Eq.~(\ref{prom}) the following bounds for the difference of the entropy are obtained
\begin{eqnarray}
\frac{1}{2 T_2}\left(\omega (t') \coth \left( \frac{\omega (t')}{2 T_2}\right)- \frac{\omega^2(t')+\omega^2(t)}{2 \omega (t)} \coth\left( \frac{\omega (t)}{2 T_1}\right) \right)\leq \Delta S \leq \nonumber \\
\leq \frac{1}{2 T_1}\left(\frac{\omega^2(t')+\omega^2(t)}{2 \omega (t')} \coth\left( \frac{\omega (t')}{2 T_2}\right)- \omega(t) \coth\left( \frac{\omega (t)}{2 T_1}\right) \right)
\end{eqnarray}

%Figure 2
\begin{figure}
\centering
\subfigure[]
{\includegraphics[scale=0.3]{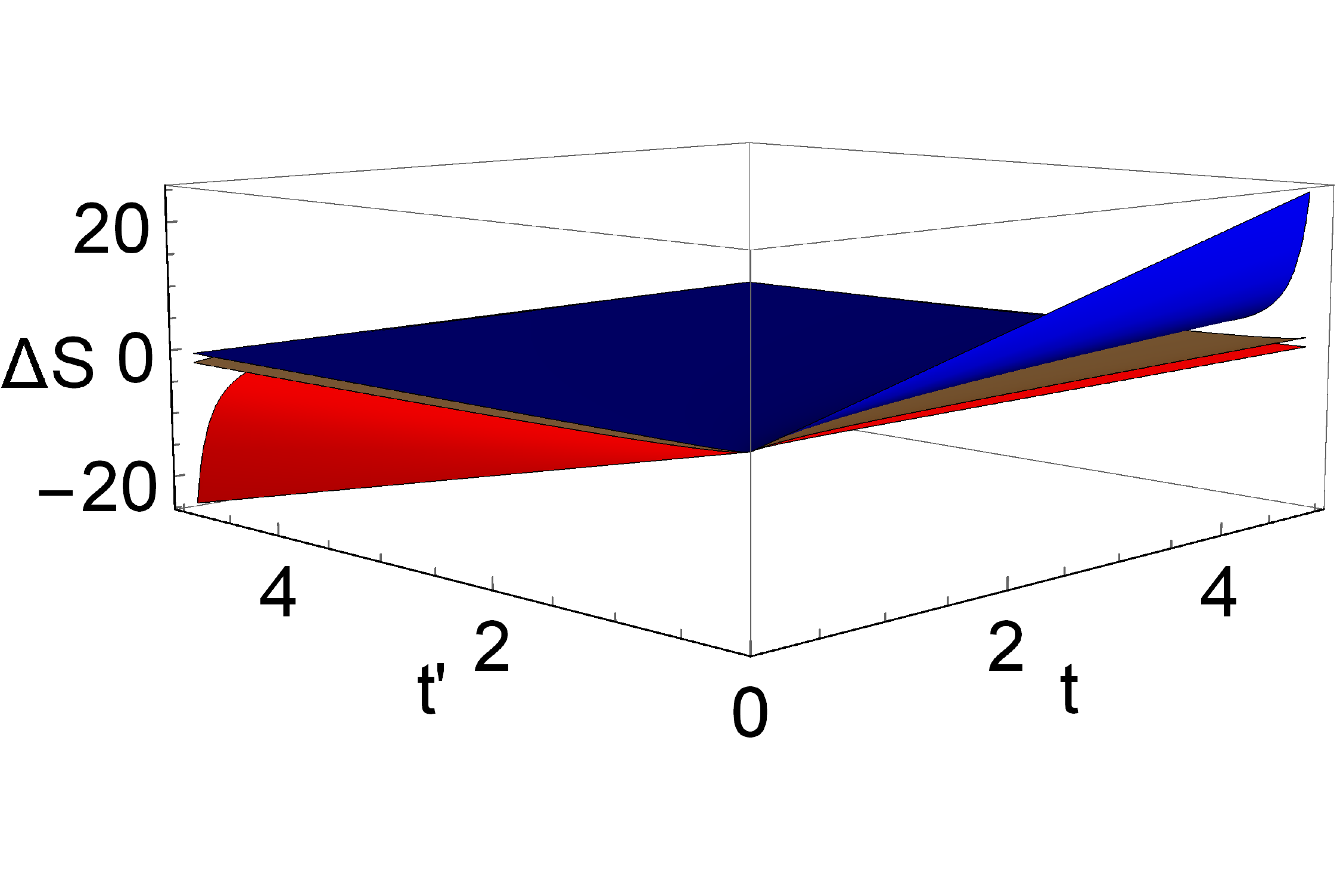} \label{fig:osci:a}} \
\subfigure[]
{\includegraphics[scale=0.27]{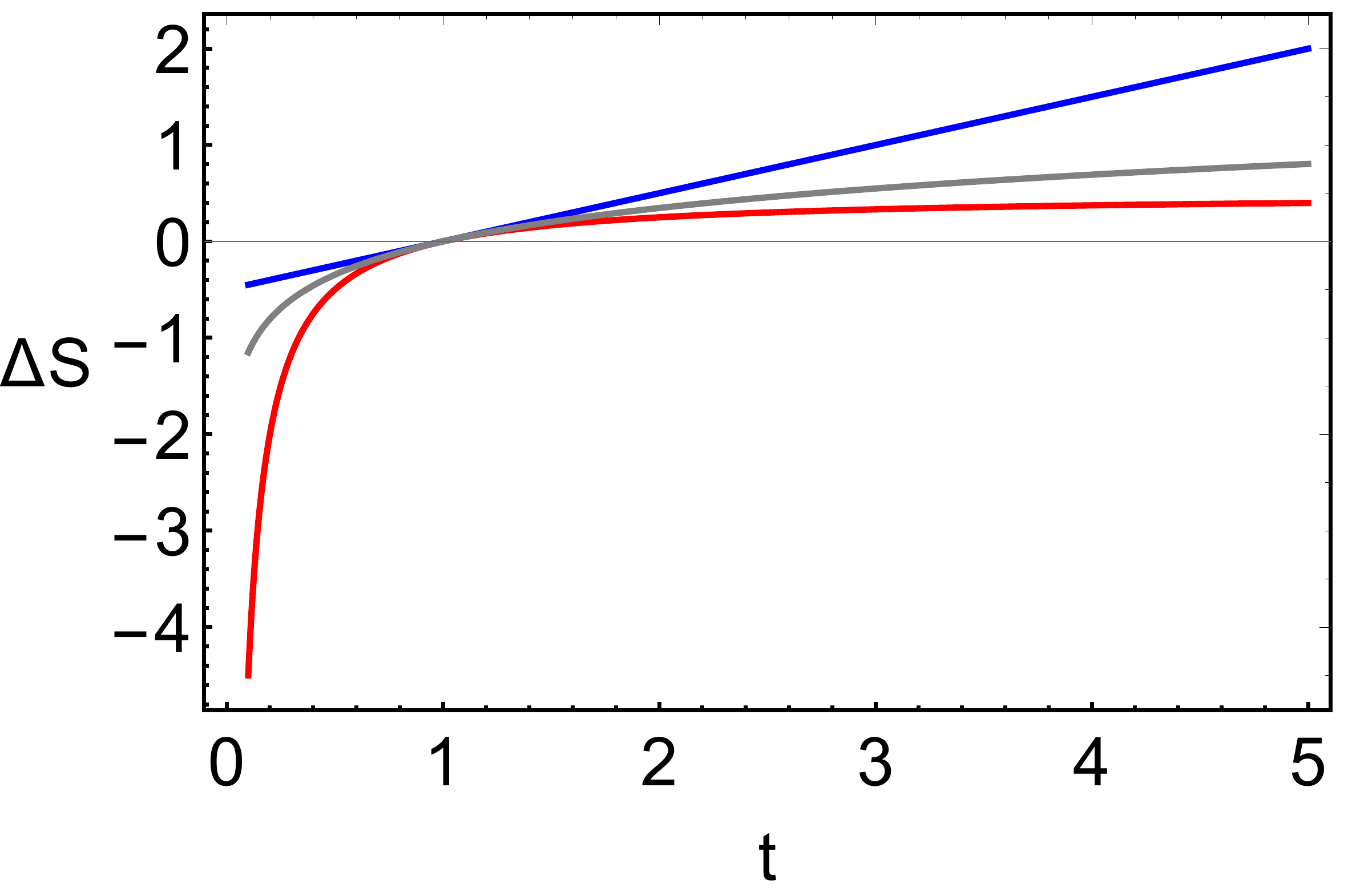} \label{fig:osci:b}} \
\caption{(a) Upper (blue), lower (red) bounds and analytical expression (gray) for the difference of
entropies $S_2(T_2,t')-S_1(T_1, t)$ for a harmonic
oscillator with time dependent frequency $\omega
(t)=\sqrt{t}$, with equal temperatures $T_1=T_2=10$. (b) a 2D cut of the previous plot for $t'=1$.}
\end{figure}

To exemplify the use of the previous results, we analyze two
different cases for $\omega (t)$. The case $\omega (t)=\omega_0
\sqrt{1+\eta \, t}$ has been studied in \cite{agarwal}, where the
authors demonstrate the presence of nonclassical effects of light as
squeezing and antibunching. The other example corresponds to the
case where $\omega (t)=\omega_0\sqrt{1+\eta \, \cos(\Omega t)}$,
which can describe the electromagnetic field inside a Paul trap and
was first studied in \cite{paul}. Additionally to these examples, we
discuss the case where the harmonic oscillator Hamiltonian is
constructed using the time dependent invariant operators $\hat{H}=\omega (t) \left(
\hat{A}^\dagger (t)
\hat{A}(t) +1/2\right)$, since the eigenfunctions of this operator
are known.

In the case of $\omega (t)=\omega_0 \sqrt{1+\eta\, t}$, the
solutions to the equation $\ddot{\epsilon}(t)+\omega^2 (t) \epsilon
(t)=0$ are the Airy functions and their derivatives.  In
fig.~\ref{fig:osci:a}, the time dependence of the bounds for $\Delta
S$ are shown for a system where $\omega (t)=\sqrt{t}$ for fixed
temperatures $T_1=T_2=10$. In this plot, the gray function
corresponds to the analytic solution of $S_2(T_2,t')
-S_1(T_1,t)$. In fig.~\ref{fig:osci:b}, one can see that there is a region ($t<t'=1$) where both the analytical and the limits for $\Delta S$ are negative and some other region ($t>t'=1$) where these quantities are positive.

\begin{figure}
\centering
\subfigure[]{
\includegraphics[scale=0.3]{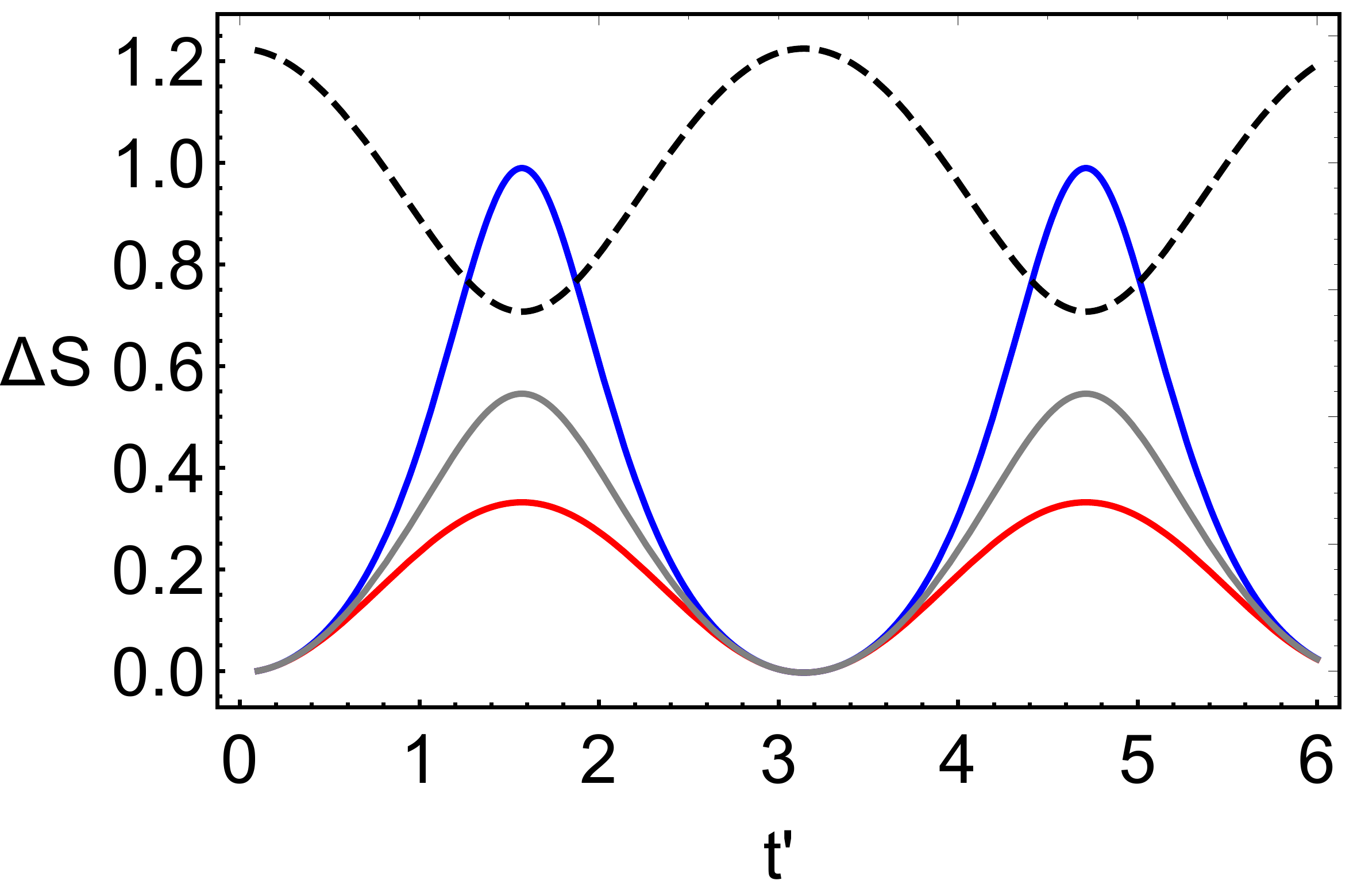}}
\subfigure[]{
\includegraphics[scale=0.3]{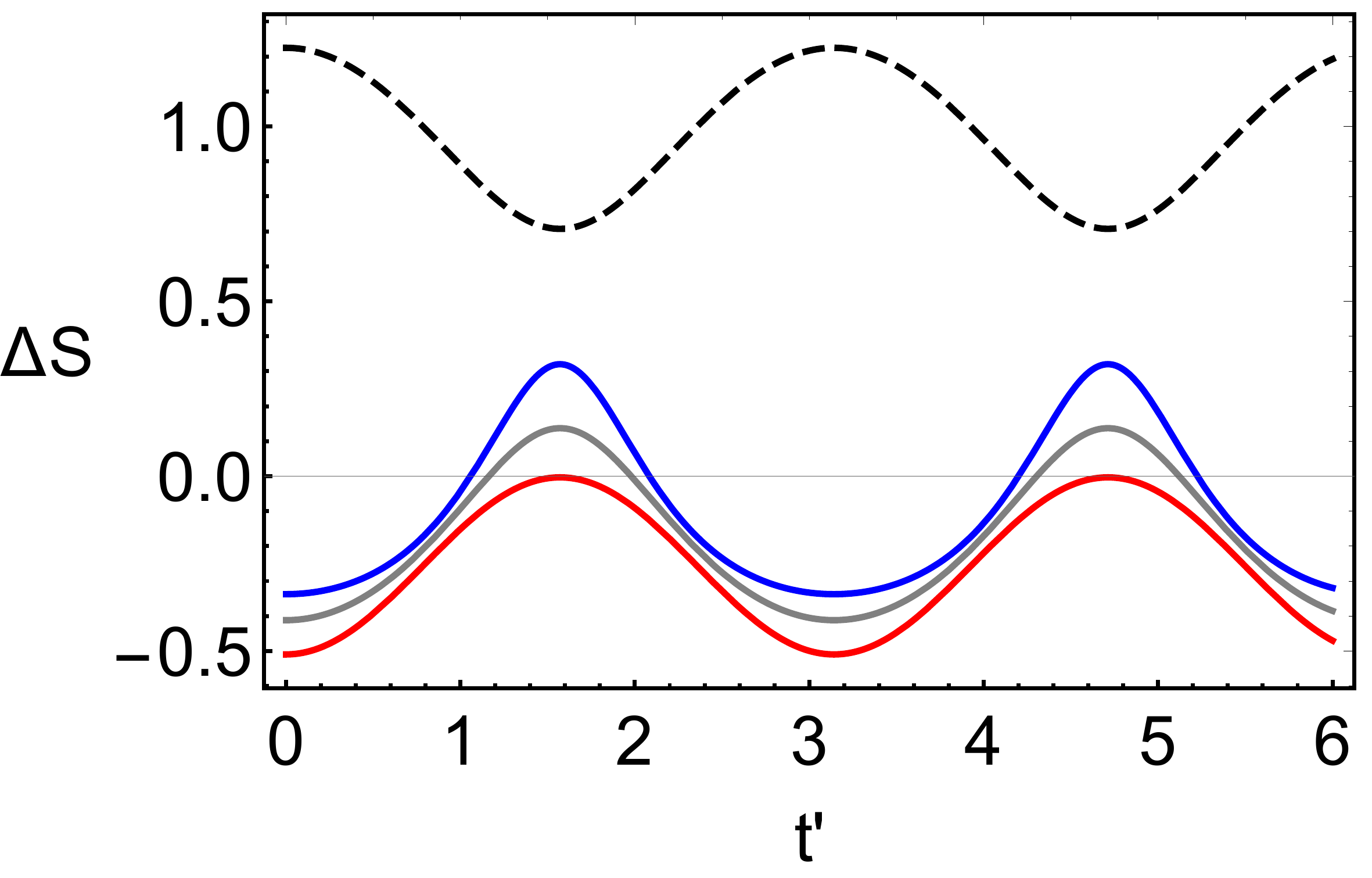}}
\caption{Upper (blue) and lower (red) bounds for the difference of
the entropies $\Delta S(t,t')$  as a function of $t'$ for a
time-dependent frequency harmonic oscillator ($\omega
(t)=\sqrt{1+\cos (2t)/2}$) between two thermal equilibrium states
with temperatures (a) $T_1=T_2=10$  at the fixed time $t=0.1$ and
(b) $T_1=15$, $T_2=10$  at the fixed time $t=3$. The dashed curve
corresponds to the plot of the frequency. \label{fig:paul}}
\end{figure}

When the frequency has an oscillatory dependence on time, as the one
observed in the Paul traps with $\omega (t)=\omega_0\sqrt{1+\eta
\, \cos(\Omega t)}$, the solution to the equation
$\ddot{\epsilon}(t)+\omega^2 (t) \epsilon (t)=0$ is given by the
Mathieu functions and their derivatives. In fig.~\ref{fig:paul}, the
dependence of $\Delta S$ in terms of time $t'$ is shown for two
cases $T_1=T_2$ and $T_1> T_2$ for fixed time $t=0.1$ and the
explicit frequency $\omega (t)=\sqrt{1+\cos(2 t)/2}$. In both cases,
the minimum values of the difference of the entropy occurs when the
frequency (dashed curve) has a maximum and a maximum value
corresponds to the case where the frequency has a minimum. Also it
is worth noticing that the minimal difference between the limits and
the analytic solution occurs when $\Delta S$ is minimal and has a
maximum when $\Delta S$ also has a maximum.

%Figure 3
%
%%%%%%%%%%%%%%%%%%%%%%%%%%%%%%%%%%%%%%%%

When the Hamiltonian of the system is given by the time dependent
invariants
\begin{equation}
\hat{H}(t)=\frac{1}{2}(\hat{P}^2(t)+\omega^2 (t) \hat{Q}^2 (t)) \, ,
\end{equation}
with the operators $\hat{Q}(t)=(\hat{A}+
\hat{A}^\dagger)/\sqrt{2 \omega (t)}$ and $\hat{P}(t)=i\sqrt{\omega (t) /2}(\hat{A}^\dagger- \hat{A})
$ expressed in terms of the integrals of
motion. This Hamiltonian can be interpreted as a degenerated parametric amplifier in the standard bosonic operators with time dependent frequency i.e., $\hat{H}(t)=\nu (t)(\hat{a}^\dagger \hat{a}+1/2)-(g^*(t)\hat{a}^{\dagger2}+g (t) \hat{a}^2)$. 

Using the  eigenfunctions of the operators $\hat{A}^\dagger(t) \hat{A}(t)$ and $\hat{A}^\dagger(t') \hat{A}(t')$, one has
\begin{equation}
\textrm{Tr}(e^{- \beta \hat{H}(t')} \hat{H}(t)) =\omega (t)
 \sum_{n,m=0}^\infty e^{-\beta \, \omega (t') (m+\frac{1}{2})}
 \left(n +\frac{1}{2}\right) \int dx\, dx' \, \phi_n (x,t)
 \phi_m^*(x,t') \phi_n^* (x',t) \phi_m (x',t')\,,
\label{eq:sum}
\end{equation}
where both sums over $n$ and $m$ can be done separately before
the integration using the Mehler formula
\begin{equation}
\sum_{j=0}^\infty \frac{(\tilde{\eta}/2)^j}{j!} H_j (y) H_j(y')=\frac{1}
{(1-\tilde{\eta}^2)^{1/2}} \exp \left(\frac{2 \tilde{\eta} y y'-
(y^2+y'^2)\tilde{\eta}^2}{1-\tilde{\eta}^2} \right).
\end{equation}
The sum over $m$ gives
\begin{eqnarray}
&&\sum_{m=0}^\infty e^{-\beta \omega (t')(m+{1}/{2})}
\phi_m^* (x,t') \phi_m (x',t')=\frac{e^{-\beta
\omega (t')}/{2}}{\pi^{1/2} \vert b(t') \vert
(1-e^{-2 \beta \omega (t')})^{1/2}}
\nonumber\\
&&\times
 \exp\left\{\frac{2
e^{- \beta \omega (t')} q q'- (q^2+q'^2)e^{-2 \beta
\omega (t')}}{1-e^{-2 \beta \omega (t')}} \right\}
\exp\left\{-i\frac{\dot{\epsilon}^*(t')}{2 \, \epsilon^*(t')}x^2\right\}
\exp\left\{i\frac{\dot{\epsilon}(t')}{2 \,
\epsilon(t')}x'^2\right\},
\label{eq:sum_m}
\end{eqnarray}
with $q=x/\vert \epsilon(t') \vert$ and $q'=x'/\vert \epsilon(t') \vert$.
While the sum over $n$ is equal to
\begin{equation}
\omega (t) \sum_{n=0}^\infty \left( n +\frac{1}{2}\right)
\phi_n^* (x',t) \phi_n (x,t)=\omega (t)
e^{i \frac{\dot{\epsilon}(t)}{2 \epsilon(t)}x^2} \left(\frac{1}{2}
 + x \frac{\partial }{\partial x}-\frac{\vert \epsilon(t) \vert^2}{2}
 \frac{\partial^2 }{\partial x^2}\right)e^{-i \frac{\dot{\epsilon}(t)}
 {2 \epsilon(t)}x^2} \delta(x-x');
\label{eq:sum_n}
\end{equation}
to obtain this expression, the derivative of the Hermite polynomials
$\frac {d H_{n+1} (x)}{dx} = 2 (n+1) H_n(x)$ together with the
recursion relation $H_{n+1} (x)=2x H_n (x)-\frac{d H_n (x)}{dx}$
were used. Substituting Eqs.~(\ref{eq:sum_m}) and (\ref{eq:sum_n})
into (\ref{eq:sum}), one arrives at the following expression:
\[
\textrm{Tr}(e^{-\beta \hat{H}(t')} \hat{H}(t))= \frac{\omega(t)
e^{-\beta \omega (t')/2}}{4(1-e^{-2 \beta \omega (t')})^{1/2}
\left(\tanh\left({\beta \omega (t')}/{2}\right) \right)^{3/2}}
f(t,t') \, , \nonumber \\
\]
with the definition
\[
f(t,t')=  \vert \epsilon (t) \vert^2  \vert \dot{\epsilon}(t')
\vert^2+\vert \dot{\epsilon}(t)\vert^2 \vert \epsilon (t') \vert^2- 2
\, \textrm{Re}\,(\epsilon(t) \dot{\epsilon}^*(t))\, \textrm{Re}\,
(\epsilon(t') \dot{\epsilon}^*(t')) \, .
\]
Notice that the function $f(t,t')$ is symmetrical under the
interchange of $t$ and $t'$, so the same expression can be used to
obtain $\textrm{Tr}(e^{-\beta \hat{H}(t)} \hat{H}(t'))$.

The previous expression yields to the result
\begin{equation}
\frac{1}{Z(T, \hat{H}(t'))}\textrm{Tr}(e^{-\beta \hat{H}(t')}
\hat{H}(t))=\frac{\omega (t)}{4} \coth\left( \frac{ \omega (t')}{2 T}\right)
f(t,t') \, ,
\end{equation}
while the mean value of the energy is
\begin{equation}
E(T, t)=({\omega (t)}/{2})\,\coth \left({ \omega (t)}/{2 T}\right)
\, .
\end{equation}
From these expressions, one can see that the difference of the
entropies $S(T_2,
t')-S(T_1, t)$  in the system at times $t$ and $t'$ has the following bounds:
\begin{eqnarray}
\frac{\omega (t')}{2 T_2} \left(\coth \left( \frac{\beta_2
\omega (t')}{2}\right)-\frac{1}{2} \coth \left(
\frac{\beta_1 \omega (t)}{2}\right) f(t, t') \right)
\leq S(T_2, t')- S(T_1, t) \leq \nonumber \\
\leq \frac{\omega (t)}{2 T_1} \left( \frac{1}{2}
\coth \left( \frac{\beta_2 \omega (t')}{2}\right)f(t,t')
- \coth \left( \frac{\beta_1 \omega (t)}{2} \right) \right) \, .
\end{eqnarray}
While the exact expression of the entropies can be obtained from the expression
\begin{equation}
S(T, t)=\bar{n}(T,t) \ln \big[ ({1+\bar{n}(T, t)})/{\bar{n}(T, t)}
\big]+ \ln (\bar{n}(T, t)+1), \quad \bar{n}(T, t)=(e^{ \omega
(t)/T}-1)^{-1} \, .
\end{equation}

\begin{figure}
\centering
\subfigure[]{
\includegraphics[scale=0.3]{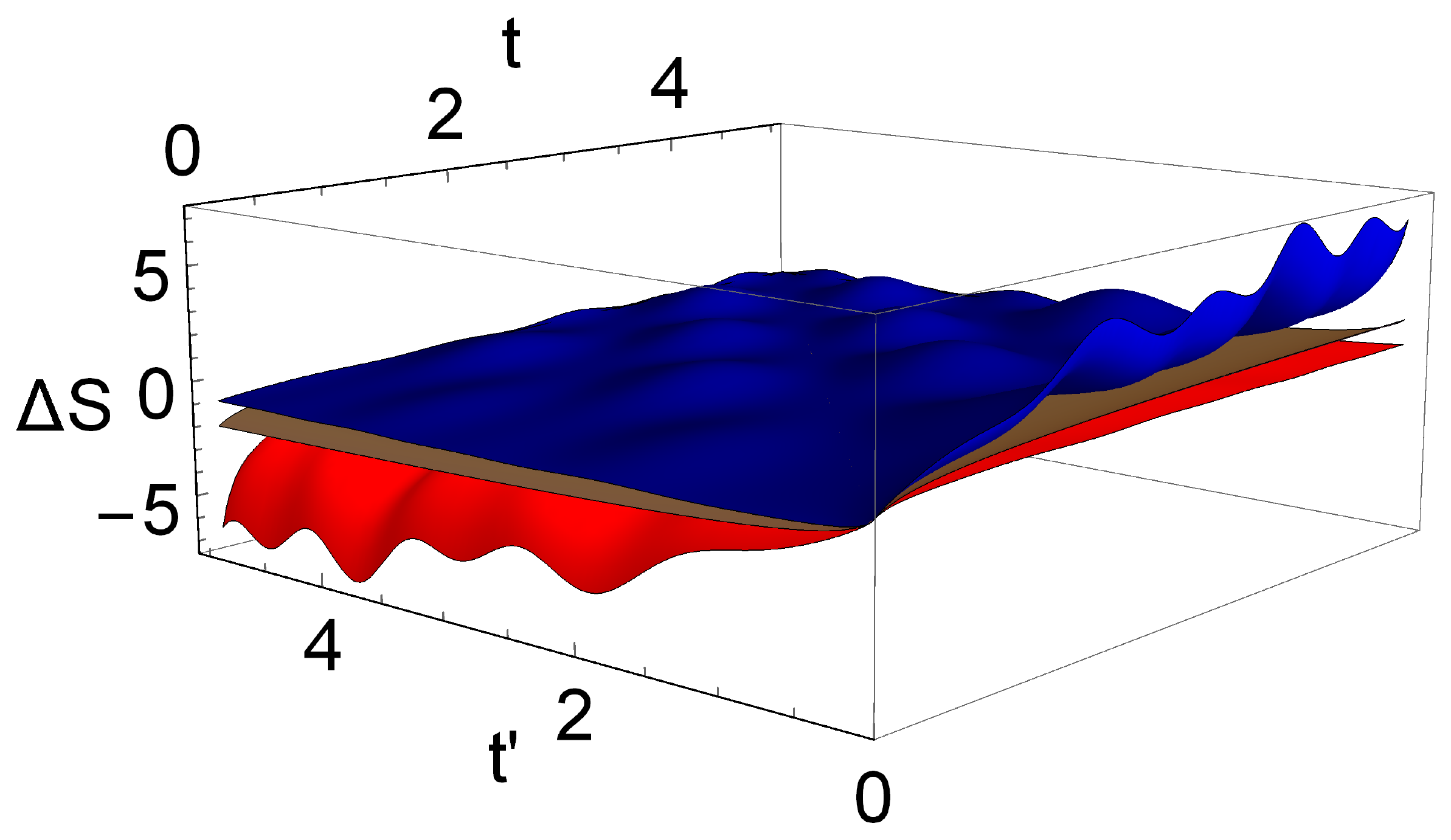}\label{fig:oscillator:a}} \
\subfigure[]{
\includegraphics[scale=0.3]{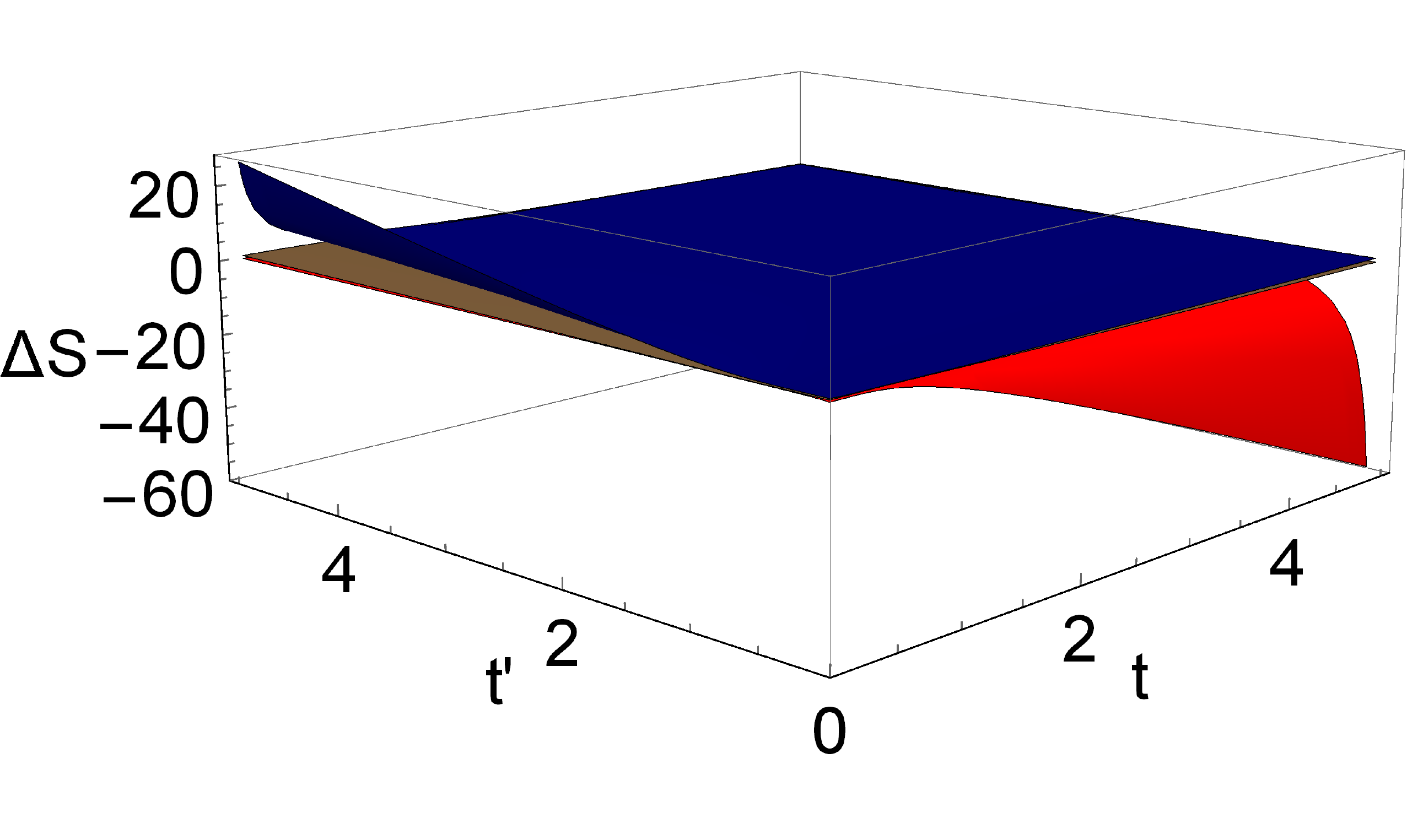}\label{fig:oscillator:b}}
\caption{Upper (blue) and lower (red) bounds for the difference of
entropies $S(T_2,t')-S(T_1, t)$ (gray) for a harmonic oscillator
with time-dependent frequency $\omega (t)=\sqrt{t}$. Here, (a)
$T_1=T_2=10$ and (b) $t=5$ and $t'=10$. }
\end{figure}

In fig.~\ref{fig:oscillator:a}, the upper and lower bounds are
plotted in terms of times $t$ and $t'$, and in
fig.~\ref{fig:oscillator:b} the dependence of these bounds in terms
of $T_1$ and $T_2$ is shown. One can see a small variation in the
time dependence and a very steady behavior in terms of temperatures.

%Figure 4

\section{Summary and concluding remarks}
When a thermal equilibrium system interacts with an external source,
either by interchanging particles or only energy, its thermodynamic
quantities as the entropy, internal energy and Helmholtz and
Gibbs potentials present a change that can be sudden or not
depending on the type of interaction with the environment.  The main results of our work are the following: we demonstrated that the change in these quantities has
upper and lower bounds when the system achieves thermal equilibrium
after the interaction. Using the relative entropy between two
thermal equilibrium states,  the upper and lower
bounds of the difference of entropies $\Delta S$ and the Helmholtz and
Gibbs potentials were obtained. In the case where the thermal equilibrium states
are expressed in terms of the Hamiltonian eigenvectors, these bounds
can be written as a sum of the Franck--Condon factors of the
two systems. From this, our results can be of interest in the measurement of the vibronic structure of electronic lines in molecules. The possibility of use this bounds to approximate the
analytic values is also discussed as the limits can be obtained
through the calculation of mean values of the Hamiltonians and the number operator before and after a sudden interaction between the
system and an environment.

As examples of applications of the general theory, the bounds for
the difference of the entropy are studied for an arbitrary qubit
system. In this case, we showed that the bounds have a minimal
difference when the Bloch vectors of the Hamiltonians from the
initial and final equilibrium states are parallel and have a maximal
difference when the Bloch vectors are antiparallel. Also it is
noticed that as both states tend to the most mixed density matrix
($\mathbf{I}/2$) as the temperature increases, then this difference
decreases independently of the Hamiltonians.

Also the limits for $\Delta S$ were obtained for a harmonic
oscillator with time dependent frequency. The bounds were calculated
using the eigenvalues and eigenvectors of the constants of motion of the
system for two particular cases: $\omega (t)=\omega_0\sqrt{t}$, which has been used to show nonclassical properties, and
$\omega (t)=\omega_0 \sqrt{1+\eta \cos (\Omega t)}$, which
describes an electromagnetic field in a Paul trap. These limits were
also calculated for a harmonic oscillator Hamiltonian written in
terms of the operators $\hat A (t)$ and $\hat A^\dagger
(t)$. The results obtained for these systems can be applied to the different potentials that can be approximated by a harmonic oscillator.

\section*{Acknowledgments}
This work was partially supported by CONACyT-M\'exico under Project
No.~238494. The work of V. I.~Man'ko and J. A.~L\'opez-Sald\'ivar was
performed at the Moscow Institute of Physics and Technology, where
V.I.~Man'ko was partially supported by the Russian Science
Foundation under Project No.~16-11-00084. Also V. I.~Man'ko
acknowledges the partial support of the Tomsk State University
Competitiveness Improvement Program.

\appendix
\section{Harmonic oscillator}
To calculate the mean value of the Hamiltonian at time $t$ with
respect of the density matrix at time $t'$, we use the SU(1,1)
algebra decomposition of the Hamiltonian given in Eq. (\ref{hal}).
The SU(1,1) generators are $\hat{K}_+(t)=\hat{A}^\dagger (t)/2$,
$\hat{K}_-(t)=\hat{A}^2(t)/2 $, and $\hat{K}_0 (t) =(\hat{A}^\dagger
(t) \hat{A}(t)+1/2)/2$. This decomposition allows us to write the
exponential operator $e^{-\beta \hat{H}(t')}$ as the product of the
elements of the algebra $e^{-\beta \hat{H}(t')}=e^{A_+(t',t,T)
\hat{K}_+}e^{\ln (A_0(t',t,T)) \hat{K}_0} e^{A_-
(t',t,T)\hat{K}_-}$, where $A_+(t',t,T)$ and $A_0(t',t,T)$ are
\begin{eqnarray}
A_0(t',t,T)&=&\frac{4 \omega^2 (t')}{(2 \omega (t')
\cosh (\omega (t')/T)+\gamma(t,t')\sinh(\omega (t') /T))^2}\, , \nonumber \\
A_+ (t',t,T)&=&- \frac{2 \alpha^* (t,t')\sinh(\omega (t')/T)}{2 \omega (t')
\cosh (\omega (t')/T)+\gamma(t,t') \sinh(\omega (t') /T)} \, ,
\end{eqnarray}
and $A_-(t',t,T)=A_+^*(t',t,T)$. With this, the partition function of the system can be evaluated using the eigenstates of the operator $\hat{A}^\dagger(t) \hat{A}(t)$ as follows:
\[
{\rm Tr}(e^{-\beta \hat{H}(t')})=\sum_{l,m,n=0}^\infty c_{l,m}
\langle n,t \vert \hat{A}^{\dagger 2m} e^{\ln A_0(t',t,T)(\hat{A}^\dagger
(t)\hat{A}(t)+1/2)/2} \hat{A}^{2l}\vert n,t \rangle \, .
\]
where a Taylor expansion of the exponential for $\hat{A}^2(t)$ and
$\hat{A}^{\dagger 2}(t)$ was performed, and the coefficients $c_{l,m}$ are given by
\[
c_{l,m}=\left( \frac{A_+ (t',t,T)}{2}\right)^m
\left( \frac{A_+^* (t',t,T)}{2}\right)^l \frac{1}{m!\, l!} \, .
\]
This sum can be rewritten as
\[
{\rm Tr}(e^{-\beta \hat{H}(t')})=A_0^{1/4} (t',t,T)
\sum_{n,m=0}^\infty  \frac{A_0^{n/2}(t',t,T)n!}{(m!)^2
(n-2m)!} \left( \frac{\vert A_+(t',t,T)\vert^2}{4 A_0(t',t,T)}\right)^m \, ,
\]
this infinite sum can be truncated for values where the factorial
$(n-2m)!< 0$ ($m>n/2$). So the previous equation can be expressed in terms
of the Legendre polynomials $P_n (z)$
\[
{\rm Tr}(e^{-\beta \hat{H}(t')})=A_0^{1/4}(t',t,T)
\sum_{n=0}   A_0^{n/2}(1-x)^{n/2}P_n
\left(\frac{1}{\sqrt{1-x}}\right) \, ,
\]
with $x=\frac{\vert A_+ (t',t,T) \vert^2}{A_0 (t',t,T)}$. Finally, using the generating function of the Legendre polynomials
the following result is obtained:
\begin{equation}
Z(\hat{H}(t'),T)=\frac{A_0^{1/4}(t',t,T)}{(1-2 A_0^{1/2}(t',t,T)
+A_0(t',t,T)-\vert A_+ (t',t,T) \vert^2)^{1/2}} \, ,
\label{zeta}
\end{equation}
using the properties of the classical solutions $\dot{\epsilon}(t) \epsilon^* (t) -\epsilon(t) \dot{\epsilon}^*
(t)=2i$, it can be seen that the partition function gives the standard result
\begin{equation}
Z(\hat{H}(t'),T)= \frac{1}{2 \sinh \left(\frac{\omega (t')}{2T}\right)}  \, .
\end{equation}
The mean values of the operators $\hat{K}_0(t)$ and $\hat{K}_\pm(t)$ can be calculated by differentiating Eq.~(\ref{zeta}) with respect to the functions $\ln (A_0(t',t,T))$, $A_+(t',t,T)$, and $A^* (t',t,T)$, respectively. This procedure gives the following expressions
\begin{equation}
\frac{1}{Z(\hat{H}(t'),T)}{\rm Tr}(e^{-\beta \hat{H}(t')}\hat{K}_0(t))
=\frac{1-A_0(t',t,T)+\vert A_+(t',t,T) \vert^2}{4(1-2 A_0^{1/2}(t',t,T)
+A_0(t',t,T)-\vert A_+(t',t,T)\vert^2)} \, ,
\label{k0}
\end{equation}
and
\begin{eqnarray}
\frac{1}{Z(\hat{H}(t'),T)}{\rm Tr}(e^{-\beta \hat{H}(t')}\hat{K}_+(t))
=\frac{A_+^*(t',t,T)}{2(1-2 A_0^{1/2}(t',t,T)+A_0(t',t,T)
-\vert A_+(t',t,T)\vert^2)} \, , \nonumber \\
\frac{1}{Z(\hat{H}(t'),T)}{\rm Tr}(e^{-\beta \hat{H}(t')}\hat{K}_
-(t))=\frac{A_+(t',t,T)}{2(1-2 A_0^{1/2}(t',t,T)+A_0(t',t,T)
-\vert A_+(t',t,T)\vert^2)} \, ,
\label{kmas}
\end{eqnarray}
then substituting  Eqs.~(\ref{k0}) and (\ref{kmas}) into
(\ref{hal}) and using the property $\dot{\epsilon}(t) \epsilon^* (t) -\epsilon(t) \dot{\epsilon}^*
(t)=2i$, we obtain the mean value of $\hat{H}(t)$ in the state $e^{-\beta\hat{H}(t')}/Z(\hat{H}(t'),T)$ given in
Eq.~(\ref{prom}).

\bibliographystyle{aipnum4-1}
%\bibliography{newb}

%

\end{document}